\documentclass[aps,prd,floatfixy,superscriptaddress,showpacs,showkeys]{revtex4}
\usepackage{amssymb}
\usepackage{amsmath}
\usepackage{amsfonts}
\usepackage{epsfig}
\usepackage{graphicx}
\usepackage{tabularx}
\usepackage{latexsym}
\usepackage{subfigure}
\usepackage{verbatim}
\usepackage{color}
\usepackage{braket}
\usepackage{multirow}
\usepackage{dcolumn}
\usepackage{bm}
\newcommand{\seq}{\begin{subequations}}
\newcommand{\sen}{\end{subequations}}
\newcommand{\eq}{\begin{eqnarray}}
\newcommand{\en}{\end{eqnarray}}

\def\ds{D^{\ast  0}}

\def\L2{\Lambda^2}

\def\eq{\begin{eqnarray}}
\def\en{\end{eqnarray}}

\usepackage[normalem]{ulem}

\renewcommand\sout{\bgroup \color{red} \ULdepth=-.5ex \ULset}

\def\ds{D^{\ast  0}}

\def\L2{\Lambda^2}

\def\eq{\begin{eqnarray}}
\def\en{\end{eqnarray}}

\def\L{{\cal L}}

\def\ds{d^*}
\def\vk{\vec{k}}
\def\vr{\vec{\rho}}
\def\vl{\vec{\lambda}}

\begin{document}

\title{\boldmath Decay width of $d^*(2380)
\to NN \pi\pi$ processes  }

\author{Yubing Dong}
\affiliation{Institute of High Energy Physics, Chinese Academy of
Sciences, Beijing 100049, China} \affiliation{Theoretical Physics
Center for Science Facilities (TPCSF), CAS, Beijing 100049, China}

\author{Fei Huang}
\affiliation{School of Physical Sciences, University of Chinese
Academy of Sciences, Beijing 101408, China}

\author{Pengnian Shen}
\affiliation{College of Physics and Technology, Guangxi Normal
University, Guilin  541004, China} \affiliation{Institute of High
Energy Physics, Chinese Academy of Sciences, Beijing 100049, China}
\affiliation{Theoretical Physics Center for Science Facilities
(TPCSF), CAS, Beijing 100049, China}

\author{Zongye Zhang}
\affiliation{Institute of High Energy Physics, Chinese Academy of
Sciences, Beijing 100049, China} \affiliation{Theoretical Physics
Center for Science Facilities (TPCSF), CAS, Beijing 100049, China}
\date{\today}

\begin{abstract}

The decay widths of four-body double-pion decays $\ds\to pn
\pi^0\pi^0$, $\ds\to pn \pi^+\pi^-$, and iso-scalar parts of $\ds\to
pp \pi^0\pi^-$ and $\ds\to nn \pi^+\pi^0$ are explicitly calculated
with the help of the $d^*$ wave function obtained in a chiral SU(3)
quark model calculation. The effect of the dynamical structure on
$\ds$'s width is analyzed both in the single $\Delta\Delta$ channel
and coupled $\Delta\Delta$ and $CC$ channel approximations. It is
found that in the coupled-channel approximation, the obtained
partial decay widths of $\ds\to pn \pi^0\pi^0$, $\ds\to pn
\pi^+\pi^-$, and those of $d^*$ to the iso-scalar parts
of $pp \pi^0\pi^-$ and $nn \pi^+\pi^0$ are about $7.4$MeV,
$16.4$MeV, $3.5$MeV and $3.5$MeV, respectively As a consequence, the
total width is about $64.5$MeV. These widths are consistent with
those estimated by using the corresponding cross section data in our
previous investigation and also the observed data. But in the single
$\Delta\Delta$ channel approximation, the widths are still almost
2-times larger than the measured values. Apparently, the explicitly
calculated width together with the evaluated mass of $d^*$ in the
coupled $\Delta\Delta$ and $CC$ channel approximation can well
explain the observed data, which again supports our assertion that
the $\ds$ resonance is a six-quark dominated exotic state.
\end{abstract}



\maketitle


Since X(3872) was reported by the Belle and Babar Collaborations in
2003, many new exotic resonances were observed in Belle, Babar,
BEPCII, LHCb and many other facilities. Those new resonances, like
$X(3872), Z_c(3900)$, $Z_b(10610)$, and $Z_b(10650)$ (or
$\Lambda_c(2900)$, $p_c(4380)$, and $p_c'(4450)$) et al., cannot
simply be understood by the well-established conventional
$q$-$\bar{q}$ (or qqq) potential models. In particular, most of them
are very near to the meson-meson (or meson-baryon) thresholds with
rather narrow widths and some of them are even charged. Different
interpretations, such as pentaquark states, molecule structures,
triangle singularities, cusp and threshold effect, for those new
resonances have been proposed.

Except for the possible multi-quark states with baryon number ${\cal
B}=0,~1$ mentioned above, the study of the 6-quark states (or
dibaryons) also has a long history. In recent years, CELSIUS/WASA
and WASA@COSY Collaborations have clearly observed a resonance-like
structure in double pionic fusion channels $pn\to d\pi^0\pi^0$ and
$pn\to d\pi^+\pi^-$ when they studied the ABC effect and in dealing
with the neutron-proton scattering data with newly measured
analyzing power $A_y$. This possible resonance has a mass of about
$2380$MeV and a width of about $70$MeV.~\cite
{CELSIUS-WASA,CELSIUS-WASA1,ABC}. Because the observed resonance
cannot simply be explained by either the intermediate Roper
excitation or the t-channel $\Delta\Delta$ process, they proposed a
$d^*$ hypothesis, in which its quantum number, mass, and width are
$I(J^P)=0(3^+)$, $M \approx 2370$ MeV and $\Gamma \approx 70$
MeV~\cite{CELSIUS-WASA,Bashkanov} ( in their recent
paper~\cite{Bashkanov}, the averaged mass and width, from the
elastic scattering and two-pion productions, are $M \approx 2375$
MeV and $\Gamma \approx 75$ MeV, respectively). Since its baryon
number is 2, it would be regarded as a dibaryon, and could be either
"an exotic compact particle or a hadronic molecule"~\cite{CERN}.
Moreover, according to the experimental data, the mass of $d^*$ is
about 80~MeV smaller than the $\Delta\Delta$ threshold and about
$70~MeV$ larger than the $\Delta\pi N$ threshold, so the threshold
(or cusp) effect is expected to be not so important as that in the
systems of XYZ particle cases, and therefore, the internal structure
of $d^*$ would be essentially significant.

This newly observed $d^*$ causes a great attention of theoreticians.
In fact, the existence of the non-trivial six-quark configuration
with $I(J^P)=0(3^+)$ (called $d^*$ lately) has intensively been
studied since Dyson's estimation~\cite{Dyson}. Many phenomenological
approaches, like the group classification method~\cite{Dyson}, the
cloudy bag model~\cite{Thomas}, the quark potential
model~\cite{Oka,Wang,Yuan}, etc., have been applied to the
investigation of $d^*$'s structure in the past. The
hidden dibaryons in one- and two-pion productions in NN collisions
are also studied recently~\cite{Kukulin}. One of those calculations
reported in 1999 should be specially mentioned. In such a
calculation, a $\Delta\Delta$ channel and a hidden-color channel
(denoted by CC afterward) were taken into account simultaneously, a
binding energy of about $40-80$ MeV was predicted, which is
consistent with the recently observed $d^*$, and the importance of
the contribution from the $CC$ channel was pointed out~\cite{Yuan}.
Unfortunately, the width of the state was not calculated in that
paper.

Since COSY's discovery, there are mainly three types of
explanations. Based on the SU(2) quark model, Ref. \cite{Wang1}
considered it as a $\Delta\Delta$ resonance and performed a
multi-channel scattering calculation. They obtained a binding energy
of about 71 MeV with respect to the threshold of the $\Delta\Delta$
channel and a width of about 150 MeV where $\Gamma_{NN}=14$ MeV and
$\Gamma_{inel}=136$ MeV, which is apparently much larger than the
data. On the other hand, Ref. ~\cite{Gal} studied a three-body
system of $\Delta N \pi$ and found a resonance pole with a mass of
$2363\pm 20$MeV and a width of $65\pm 17$MeV. However, 
one mentioned that an additional factor of 2/3 should not be
included in the estimation for comparing with the observed
width~\cite{Adlarson}. An important view point, argued by
Bashkanov, Brodsky and Clement \cite{Brodsky} in 2013 is that a
dominant hidden-color structure (or six-quark configure) of $d^*$ is
necessary for understanding its strong coupling. Sooner after,
following our previous prediction~\cite{Yuan}, Huang and his
collaborators made an explicit dynamic calculation by using a chiral
SU(3) quark model in the framework of the Resonating Group Method
(RGM), with which the ground state baryon properties, the
baryon-baryon scattering and binding behaviors have been well
reproduced~\cite{Zhang1,Zhang2,Huang2}, and showed that the $\ds$
state has a mass of $2380-2414$MeV, which agrees with COSY's
observation, and does have a ``hidden-color" (CC) configuration of
about $66-68$\% in its wave function~\cite{Huang}. Based on the
obtained wave functions of $\ds$ and deuteron, Dong and his
collaborators calculated the partial decay widths of the "Golden"
decay channel (with emitted 2$\pi$) $d^*\to
d+2\pi^0(\pi^+\pi^-)$ recently~\cite{Dong}. In the
calculation, both the single $\Delta\Delta$ channel and coupled
$\Delta\Delta+CC$ channel are considered, and the dynamical effect
on the $d^*$'s width is explicitly given. It is shown that the
consideration of the hidden-color configuration inside $d^*$ could
greatly suppress its width and the resultant widths  for both
$d^*\to d\pi^0\pi^0$ and $d^*\to d\pi^+\pi^-$ are in good agreement
with the experimental measurements. Then, by making use of the
observed cross sections in other possible decay channels of $\ds$,
they gave an estimate of the total $\ds$'s width of about $69$MeV,
which is fairly good in agreement with the data~\cite{Dong}. All of
these outcomes imply that $\ds$ is probably a six-quark dominated
exotic state.

However, a blemish in our previous calculation is that the four-body
$\pi\pi$ decays $d^*\to pn\pi^0\pi^0$, $d^*\to pn\pi^+\pi^-$, and
iso-scalar parts of $d^*\to pp\pi^0\pi^-$ and $d^*\to nn\pi^+\pi^0$
were not explicitly calculated~\cite{Dong}. A naive conjecture from
the $\ds \to \Delta\Delta \to np\pi\pi$ process showed a very large
value~\cite{Brodsky}, which does not fit the observed
data~\cite{Bashkanov}. Since in these four-body decays, the only
difference with the corresponding three-body decays in our previous
calculation is that the final proton and neutron are free particles
rather than a weekly bound state of the deuteron, it is our purpose
to check if the obtained $d^*$ wave function with our conjectured
structure of $\ds$, a six-quark dominated exotic state, can also
reproduce the partial widths for these four-body double-pion decay
processes.

The phenomenological effective Hamiltonian for the pseudo-scalar
interaction among quark, pion, and quark in the non-relativistic
approximation reads \eq {\cal H}=g_{qq\pi}\vec{\sigma}\cdot
\vk_{\pi}\tau\cdot\phi\times
\frac{1}{(2\pi)^{3/2}\sqrt{2\omega_{\pi}}}, \en where $g_{qq\pi}$ is
the coupling constant, $\phi$ stands for the $\pi$ meson field,
$\omega_{\pi}$ and $\vk_{\pi}$ are the energy and three-momentum of
the $\pi$ meson, respectively, and ${\bf \sigma} ({\bf \tau})$
represents the spin (isospin) operator of a single quark. In the
conventional constituent quark model, the wave functions are \eq
\mid N> \,=\frac{1}{\sqrt{2}}\Big
[\chi_{\rho}\psi_{\rho}+\chi_{\lambda} \psi_{\lambda}\Big
]\Phi_N(\vr,\vl) \en for the nucleon and \eq
\mid\Delta>\,=\chi_s\psi_s\Phi_{\Delta}(\vr,\vl) \en for the
$\Delta(1232)$ resonance. In Eqs. (2-3), $\chi$ and $\psi$ stand for
their spin and isospin wave functions, $\Phi_N(\vr,\vl)$ and
$\Phi_{\Delta}$ are the spatial wave functions of the nucleon and
$\Delta$ resonance, respectively, and
$\vec{\rho}=\frac{1}{\sqrt{2}}(\vec{r}_1-\vec{r}_2)$ and
$\vec{\lambda}=\frac{1}{\sqrt{6}}(\vec{r}_1+\vec{r}_2-2\vec{r}_3)$
are the conventional Jacobi coordinates for the internal motion.
Then, in terms of the measured decay width of the $\Delta\to \pi N$
process, for example $\Gamma_{\Delta^+\to\pi^0 p}\sim 117~MeV$), one
can easily extract the coupling constant $g_{qq\pi}$ by calculating
$\Gamma_{\Delta \to N \pi}=\langle \Delta |{\cal
H}|N\rangle$~\cite{PDG} (the details can be found in
Ref. \cite{Dong}).

As mentioned in Ref.\cite{Huang,Dong}, our model wave function is
obtained by dynamically solving the bound-state RGM equation of the
six quark system in the framework of the extended chiral $SU(3)$
quark model, where the binding energy of $\ds$ is $\epsilon \approx
62$ MeV in the single $\Delta\Delta$ channel approximation and
$\epsilon \approx 84$ MeV if the CC channel is further considered,
and consequently, the mass of $\ds$ is
$M_{\ds}=2M_{\Delta}-\epsilon$. Further projecting the wave function
in the quark level onto the two-cluster wave function in the baryon
level, namely the physical state, we end up a wave function of $d^*$
\eq \label{wf1}\Psi_{d^*} &=&
[~\phi_{\Delta}(\vec{\xi}_1,\vec{\xi}_2)~\phi_{\Delta}(\vec{\xi}_4,\vec{\xi}_5)~
\chi_{\Delta\Delta}(\vec{R})~\zeta_{\Delta\Delta}~ +~
\phi_{C}(\vec{\xi}_1,\vec{\xi}_2)~\phi_{C}(\vec{\xi}_4,\vec{\xi}_5)~
\chi_{CC}(\vec{R}))~\zeta_{CC}~]_{(SI)=(30)}, \en where
$\phi_{\Delta}$, and $\phi_{C}$ denote the internal wave functions
of $\Delta$ and $C$ (color-octet particle) in the coordinate space,
$\chi_{\Delta\Delta}$ and $\chi_{CC}$ represent the relative wave
functions between $\Delta$s and $C$s (in the single $\Delta\Delta$
channel case, the $CC$ component is absent), and
$\zeta_{\Delta\Delta}$ and $\zeta_{CC}$ stand for the spin-isospin
wave functions in the $\Delta\Delta$ and $CC$ channels,
respectively~\cite{Huang}. It should be specially mentioned that in
such a wave function, normally called channel wave function, these
two components are orthogonal to each other, and the totally
anti-symmetric effect is implicitly included in the resultant
relative wave function through above mentioned two
steps~\cite{Huang}. To simplify the calculation without missing the
major character, the $D$-wave contribution in the following
calculations would be ignored due to its relatively smaller
contribution, although both the $S$- and $D$-wave functions exist in
our resultant wave functions. For convenience, the relative $S$-
wave function is expanded as \eq \chi(R)=\sum^4_{i\,=\,1}~c_i\,
\exp\left(-\frac{R^2}{2b_i^2}~\right). \en

In terms of obtained wave function of $d^*$, we are able to
calculate the four-body decay width of $d^*\to pn\pi^0\pi^0$ \eq
\Gamma_{d^*\to pn\pi^0\pi^0}=\frac{1}{2!2!}\int
d^3k_1d^3k_2d^3p_1(2\pi)\delta(\Delta E) \mid \overline{{\cal
M}(k_1,k_2;p_1)}\mid^2, \en where $\mid \overline{{\cal
M}(k_1,k_2;p_1)}\mid^2$ stands for the squared transition matrix
element with a sum over the final four body state and an average of
the polarizations of the initial state $\ds$, the factor of
$2!\times 2!$ is due to the identical particles of $\pi^0\pi^0$ and
$pn$, respectively, and $\delta(\Delta E)$ denotes the energy
conservation with $\Delta
E=M_{d^*}-\omega_{\pi}(k_1)-\omega_{\pi}(k_2)-E_N(p_1)-E_N(-p_1-k_1-k_2)$,
$\omega_{\pi}$ and $E_{N}$ being the the energies of the outgoing
pion and nucleon, respectively.

In the above equation, the matrix element includes the contributions
from four sub-diagrams plotted in Fig. 1.

\begin{figure}[htbp]
\centering \includegraphics [width=7cm, height=8cm]{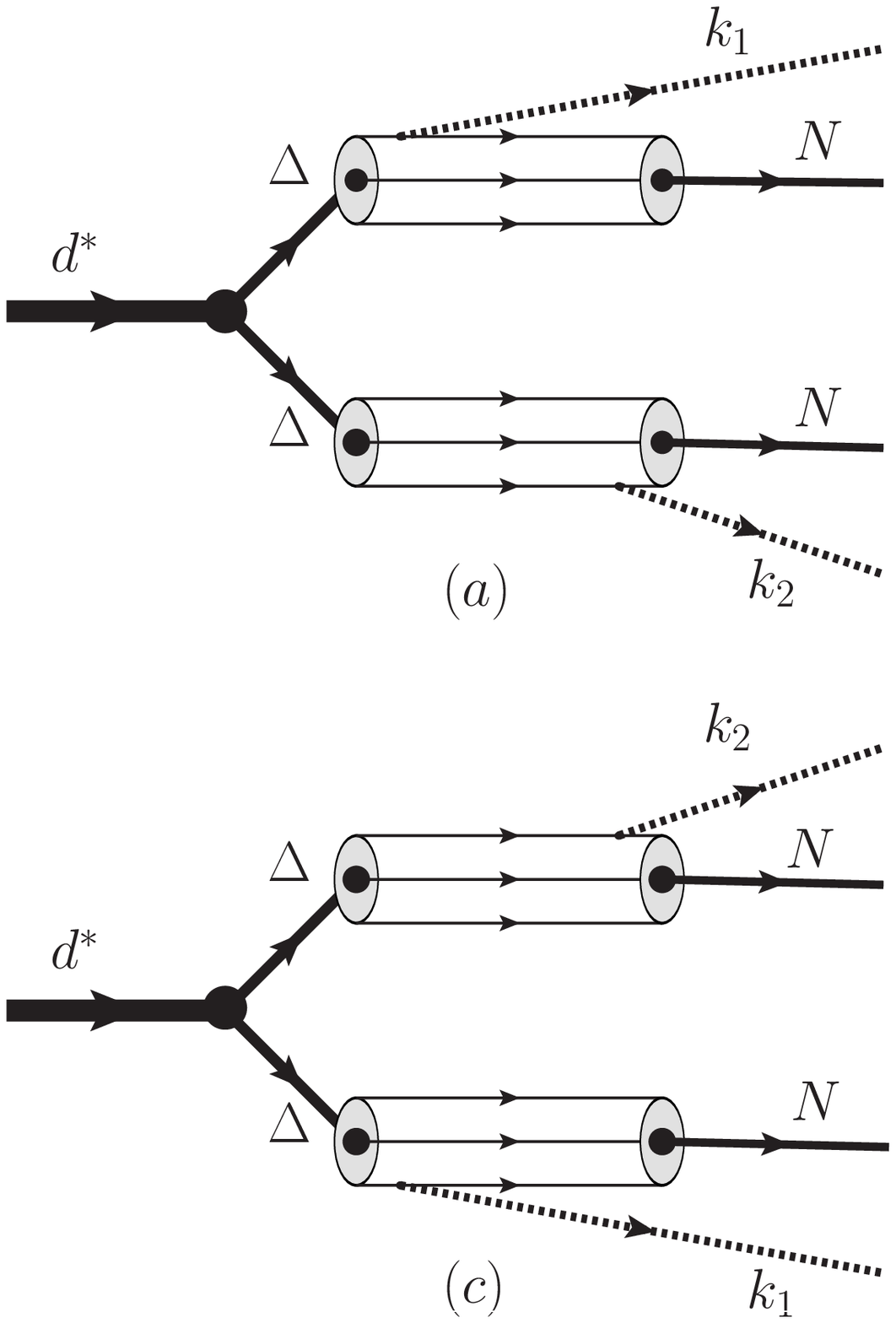}
{\hskip 1cm}
\includegraphics [width=7cm, height=8cm]{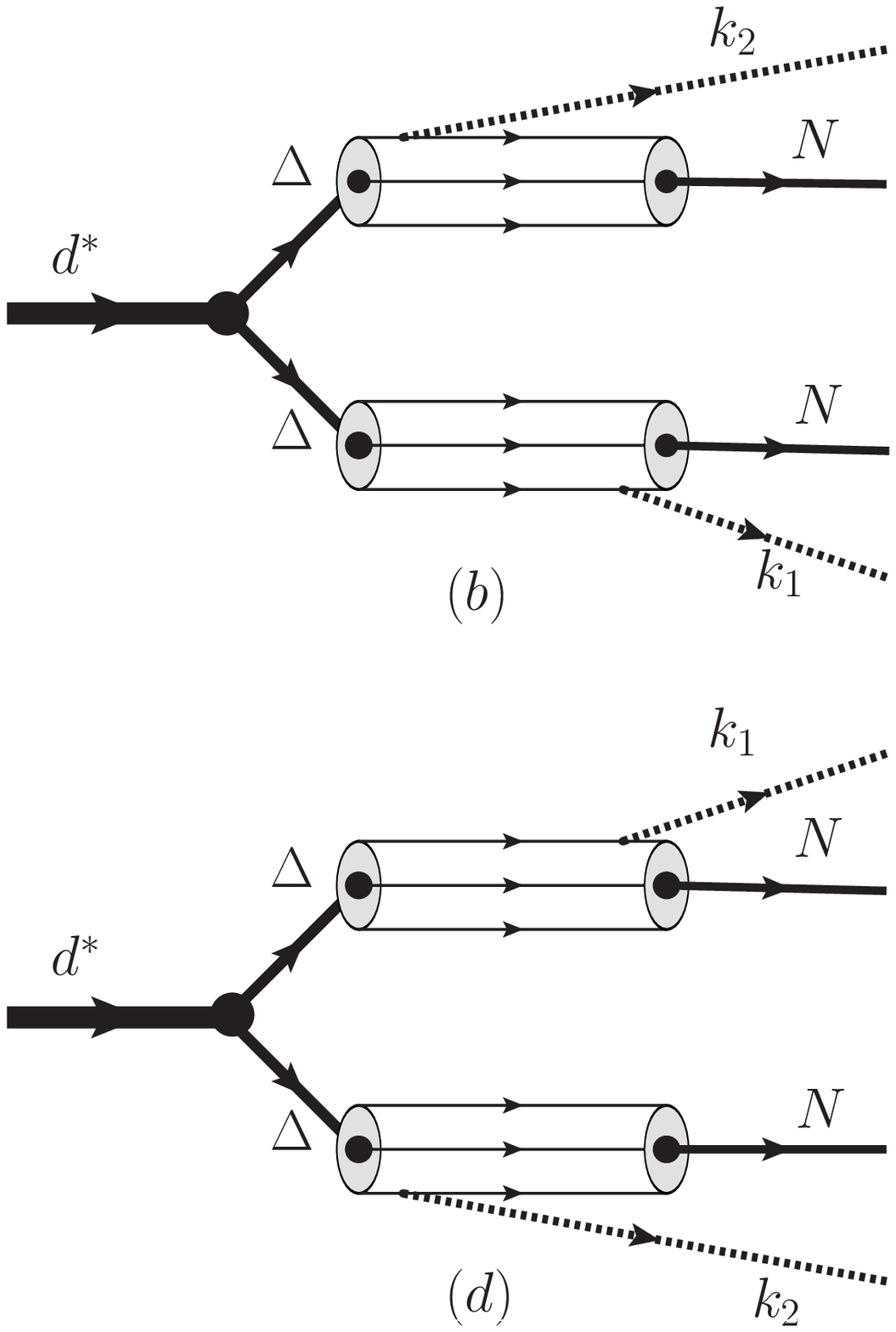}
\caption{Four possible emission ways in the decay of the $\ds$
resonance composed of the $\Delta\Delta$ structure only. Two pions
with momenta of $\vec{k}_{1,2}$ are emitted from one of the three
quarks in 2 $\Delta$s, respectively.}
\end{figure}
\noindent The explicit expression, for example Fig. 1a, reads \eq
{\cal M}^a(k_1,k_2;p_1)&=&\int d^3p_2d^3q\Big [{\cal H} {\cal
S}_f{\cal H}\Big ]\Psi_{d^*}(q)
\delta^3(\vec{p_1}+\vec{k}_1-\vec{q})\delta(\vec{p}_2+\vec{k}_2+\vec{q})\\
\nonumber &=&\int
d^3p_2\delta^3(\vec{p}_1+\vec{p}_2+\vec{k}_1+\vec{k}_2) \Big [{\cal
H}{\cal S}_f{\cal H}\Big ]\psi_{d^*}(-\vec{p}_2-\vec{k}_2), \en
where ${\cal S}_f$ is the propagator of the intermediate state.
$\Psi_{\ds}$ represents the $\ds$ wave function in the momentum
space which can be obtained by Fourier transforming the $\ds$ wave
functions in the coordinate space in both single $\Delta\Delta$
channel and coupled $\Delta\Delta+CC$ channel approximations.

In the coupled channel case, we found that there are $31.5$\%
$\Delta\Delta$ component and $68.5$\% $CC$ component in the $\ds$
wave function shown in Eq.(\ref{wf1})~\cite{Huang}. Since the pion
itself is colorless, emission of pion would not change the color
structure of the parent particle. On the other hand, the final
proton and neutron are, of course, colorless. Therefore, although
the pion can be emitted both from the colorless particle and from
the colored particle, in the leading approximation, the parent
particle should be colorless, as an estimation, the contribution
from the $CC$ component can be neglected although such a component
is the dominant component in $\ds$. In other words, the
quark-rearrangement effect is in the higher order and thus can be
ignored here. Then, the major contribution to the decay width comes
merely from the $\Delta\Delta$ component.

In the $\ds\to pn\pi^0\pi^0$ process, the obtained
partial widths are about $15.2$MeV and $7.4$MeV in the single
channel and coupled channel approximations, respectively, they are
tabulated in Tab.\ref{tab:tab1}.
\begin{table}[htbp]
\caption{Calculated partial decay widths and corresponding branching
ratios of $d^*$ in the two-body, three-body and four-body decay
channels and the total width of $d^*$. Case I and Case II denote the
single channel and coupled channel cases, respectively.}
\begin{center}
\begin{tabular}{lccccccccc}\hline\hline
 &~~~~ & \multicolumn{3}{c}{This~work} &~~~ &
 \multicolumn{2}{l}{~~~~~~~~~~~~Ref.~\cite{Dong}$^a$}&
\multicolumn{2}{l}{~~~Expt. \cite{CELSIUS-WASA1,WASA1,WASA2,WASA3,Bashkanov}}\\
\cline{3-5}
 & & Case I & \multicolumn{2}{c}{Case II} & & & & &\\
~ Wave Function & &~~~$(100\%){\Delta\Delta}$~~~ &
 \multicolumn{2}{c}{$(31.5\%)\Delta\Delta+(68.5\%)CC$}& &
 \multicolumn{2}{l}{$(31.5\%)\Delta\Delta+(68.5\%)CC$} & & \\
~~~~$M_{d^*}(MeV)$ & & 2374 & \multicolumn{2}{c}{2380} & &
\multicolumn{2}{l}{~~~~~~~~~~~~~~2380} &
\multicolumn{2}{l}{~~~~~~~~~~~~~~~~~2375} \\
\hline\hline Decay Channel & & $\Gamma(MeV)$ & ~~~~$\Gamma(MeV)$ &
$Br(\%)$ & &~~~~~$\Gamma(MeV)$ & $Br(\%)$ &
 ~~~~~~~~~$\Gamma(MeV)$ & $Br(\%)$ \\
\hline $~~d^*\to d\pi^0\pi^0~~$ & & $17.0$ & ~~$9.2$ & $14.3$ & &
~~$9.2$ & $13.3$ & ~~~~~$10.2$& $14(1)$\\
$~~d^*\to d\pi^+\pi^-~~$ & & $30.8$ & $16.8$ & $26.0$& & $16.8$
& $24.3$ & ~~~~~$16.7$& $23(2)$\\
$~~d^*\to pn\pi^0\pi^0~~$& & $15.2$ & $ ~~7.4$ & $11.5$ & & ~~$7.8$
& $11.3$  & ~~~~~~~$8.7$& $12(2)$ \\
$~~d^*\to pn\pi^+\pi^-~~$& & $33.5$ & $16.4$& $25.4$ & & $19.2$
& $27.8$ & ~~~~~~$21.8$& $30(4)$\\
$~~d^*\to pp\pi^0\pi^-~~$& & ~$7.2$ & ~~$3.5$& ~~$5.4$ & &
~~$3.9$  & ~~~$5.65$  & ~~~~~~~$4.4$& ~~$6(1)$\\
$~~d^*\to nn\pi^+\pi^0~~$& & ~$7.2$ & ~~$3.5$& ~~$5.4$ & &
~~$3.9$  & ~~~$5.65$  & ~~~~~~~$4.4$& ~~$6(1)$\\
$~~d^*\to pn~~$  &        &  ~$8.2$  & ~~$7.7$& $11.9$  & & ~~$8.3$
    & ~$12.0$  & ~~~~~~~$8.7$& $12(3)$\\
\hline
$~~Total~~$      &        & \hspace{-0.3cm} $119.1$& ~$64.5$& $99.9$
& & ~$69.1$
    & $100.0$ & ~~~~~~$74.9$& \hspace{-0.5cm}$103$\\
\hline\hline
\multicolumn{10}{l}{$^a$~Results in this column are obtained by
using the ratios of cross section data between relevant decay
channels.}
\end{tabular}
 \label{tab:tab1}
\end{center}
\end{table}

It should be mentioned that in the $\ds \to pn\pi^+\pi^-$ process,
both the ($pn$) pair and ($\pi^+\pi^-$) pair can be either
iso-scalar simultaneously or iso-vector simultaneously, namely
$[(pn)_{I_{pn}=0}(\pi^+\pi^-)_{I_{\pi\pi}=0}]_{I=0}$ or
$[(pn)_{I_{pn}=1}(\pi^+\pi^-)_{I_{\pi\pi}=1}]_{I=0}$~\cite{Bashkanov}.
According to the isospin relation, the contribution from the former
configuration ($I_{pn}=I_{\pi\pi}=0$) should be twice of that from
the $\ds \to pn\pi^0\pi^0$ configuration. But due to the isospin
violation of pion, our explicit calculation shows that the partial
widths of the first iso-scalar coupling part are $26.3$MeV and
$12.9$MeV in the single channel and coupled channel cases,
respectively, which are somewhat smaller than the expected value
from the isospin relation, just like in the $\ds \to d \pi\pi$ case.
The ratios of the iso-scalar coupling part of the charged pion decay
to the chargeless pion decay are $1.73$ and $1.74$, which is similar
to the values of $1.81$ and $1.83$ in the $\ds \to d \pi\pi$ case,
respectively. The later iso-vector coupling part would also have
some contribution. Since both components in this part have an
isospin 1 ($I_{pn}=I_{\pi_+\pi^-}=1$), its contribution would be the
same as that from the iso-scalar part of the $\ds \to pp\pi^0\pi^-$
process, which will be discussed in the next paragraph. Our
calculation gives a partial width of $7.2$MeV and $3.5$MeV for the
$d^* \to pp\pi^0\pi^-$ process in the single channel and coupled
channel case. Adding all these isospin caused effects
together, the resultant partial width of the $\ds \to pn\pi^+\pi^-$
process is about $33.5$MeV and $16.4$MeV, respectively, in the
single $\Delta\Delta$ and coupled $\Delta\Delta+CC$ channel
approximations, which are also tabulated in Tab.\ref{tab:tab1}. The
calculated partial width in the coupled channel approximation is
close to our previous estimation of $19.2$MeV~\cite{Dong} by using
the experimental data, and compatible with the experimental data of
$21.8$MeV. If we define the ratio of the partial decay widths
between the charged and chargeless double-pion decays as \eq
R=\frac{\Gamma_{\ds\to pn\pi^+\pi^-}}{\Gamma_{\ds\to pn\pi^0\pi^0}},
\en the resultant R value is about $2.20$ in the single
$\Delta\Delta$ approximation and $2.22$ in the coupled channel
approximation. Comparing with the value of $2.5$ from the isospin
relation and the similar value of $1.83$ in the $\ds \to d \pi\pi$
case, this ratio is somehow relatively smaller. This is because that
the pion isospin breaking effect caused phase space reduction in the
non-fusion double-pion production process plays a relatively weaker
role.

The partial widths of the iso-scalar parts of $\ds \to pp\pi^0\pi^-$
and $\ds \to nn\pi^+\pi^0$ processes can also be calculated in terms
of our wave function of $\ds$ in the same framework. Since these two
processes are mirror states, the widths of these decays should be
same. As shown in Tab.\ref{tab:tab1}, the calculated partial width
of $3.5$MeV in the coupled channel approximation is again compatible
with our previous estimated value of $3.9$MeV and the data of
$4.4$MeV.

From this table, one also sees that in the coupled $\Delta\Delta +
CC$ channel approximation, the total width of $d^*$ is about
$64.5$MeV, which is close to our previous estimated value of
$69.1$MeV and the observed value of $74.9$MeV. Moreover, the
calculated branching ratios for the decay processes shown in
Tab.\ref{tab:tab1} are all close to our previous estimations and are
all in acceptable ranges in comparison with the data.

In short, from this table, we have following observations:\\
$\bullet$ All the partial widths of the three-body and four-body
double-pion decays of $d^*$ are explicitly calculated in the same
framework by using our model wave function in the extended chiral
SU(3) quark model in a unified way. The obtain widths are close to
the estimations by using corresponding cross section data in our
previous paper~\cite{Dong}.\\
$\bullet$ Four-body decay width of $\ds$ in the coupled channel
approximation is much smaller than that in the single channel
approximation. This decay width suppression is mainly due to the
suppression of $\Delta\Delta$ component in the $\ds$ wave function,
and the large $CC$ component does not contribute in the lowest order
approximation. \\
$\bullet$ The partial decay width of the chargeless pion process
$\ds \to pn\pi^0\pi^0$ is close to the estimated value in our
previous paper but slightly smaller than the data due to the
approximation in the calculation. However, the corresponding
branching ratio is close to the data and our previous estimation,
because the total width is somewhat smaller than the data.\\
$\bullet$ The partial decay width of the $\ds \to pp \pi^-\pi^0$
process is also slightly smaller than the value of the data.
Consequently, the partial decay width of the $\ds \to pn \pi^+\pi^-$
process is even smaller than the data although it is still
acceptable for a crude theoretical calculation. But, the branching
ratios do not contradict the data.\\
$\bullet$ The R value shows an isospin-symmetry breaking effect due
to the mass difference between $\pi^0$ and $\pi^{\pm}$. The
explicitly calculated R value of $2.22$ in the coupled channel
approximation is slightly smaller than the value of $2.46$ extracted
in terms of the cross section data in our previous paper and the
value of $2.5$ from the theoretical isospin relation.\\
$\bullet$ The total width of $\ds$ is about $64.5$MeV, which is
still not incompatible with the value of $69.1$MeV in our previous
estimation and the data of $75$MeV.

From this calculation, one again sees that the single $\Delta\Delta$
structure cannot explain the observed data of $\ds$, but if a $CC$
component is involved, the partial decay widths of the three-body
and four-body double pion decays can be reasonably obtained and the
mass and width data of $\ds$ can be well-understood. All these
results support our assertion that one may assign the observed $d^*$
state as a $\Delta\Delta$ bound state with a dominant $CC$
component, namely the $\ds$ state is a six-quark dominated exotic
state. The real structure of $\ds$ should further be checked in
other decay modes, for instance the non-fusion triple-pion
production process~\cite{Lvqf}.

\begin{acknowledgments}

We would like to thank H. Clement and Qiang Zhao for their useful
and constructive discussions. F.Huang is grateful for
the support of the Youth Innovation Program Association, CAS. 
This work is partially supported by the National 
Natural Sciences Foundations of China under the grant Nos. 11035006, 11165005,
11475192, 11475181 11565007, the fund provided to the Sino-German
CRC 110 ``Symmetries and the Emergence of Structure in QCD" project
by the DFG, and the IHEP Innovation Fund under the No. Y4545190Y2.
\end{acknowledgments}


\begin{thebibliography}{99}

\bibitem{CELSIUS-WASA}M. Bashkanov et al., Phys. Rev. Lett. {\bf 102} 052301
(2009).

\bibitem{CELSIUS-WASA1}P. Adlarson et al., Phys. Rev. Lett. {\bf 106},
242302 (2011); P. Adlarson et al., Phys. Lett. B {\bf 721}, 229
(2013); P. Adlarson et al., Phys. Rev. Lett. {\bf 112}, 202301
(2014).

\bibitem{ABC}A. Abashian, N. E. Booth, and K. W. Crowe, Phys. Rev. Lett.
{\bf 5}, 258 (1960); N. E. Booth, A. Abashian, and K. M. Crowe,
Phys. Rev. Lett. {\bf 7}, 35 (1961); F. Plouin et al., Nucl. Phys. A {\bf 302},
413 (1978), J. Banaigs et al., Nucl. Phys. B {\bf 105}, 52 (1976).

\bibitem{Bashkanov}M. Bashkanov, H. Clement, T. Skorodko,
Eur. Phys. J. A{\bf 51}, 87 (2015).

\bibitem{CERN}{\it COSY confirms existence of six-quark states}, CERN
COURIER, July 23, 2014.

\bibitem{Dyson} J. Dyson, Phys Rev. Lett. {\bf 13}, 815 (1964).

\bibitem{Thomas} A. W. Thomas, J. Phys. G {\bf 9}, 1159 (1983).

\bibitem{Oka} M. Oka and K.Yazaki, Phys. Lett. B {\bf 90}, 41 (1980).

\bibitem{Wang} T. Goldman et al., Phys. Rev. C {\bf 39}, 1889 (1989).

\bibitem{Yuan} X. Q. Yuan et al., Phys. Rev. C {\bf 60},
045203 (1999).

\bibitem{Kukulin} M. N. Platonova and V. I. Kukulin,
Phys. Rev. C 87, 025202 (2013); Nucl. Phys. A {\bf 946}, 117 (2016).

\bibitem{Wang1} H. X. Huang et al. Phys. ReV. C {\bf 79}, 024001 (2009);
Phys. ReV. C {\bf 89}, 034001 (2014).

\bibitem{Gal} A. Gal and H. Garcilazo, Phys Rev. Lett. {\bf 111}, 172301 (2013);
Nucl. Phys. A {\bf 928}, 73 (2014); A. Gal, "Meson assisted
dibaryons", arXiv:1511.06605v1 [nucl-th].

\bibitem{Adlarson} P. Adlarson et al., Phys. Rev. C {\bf 90}, 035204
(2014).

\bibitem{Brodsky}M. Bashkanov, Stanley J. Brodsky, and H. Clement,
Phys. Lett. B {\bf 727}, 438 (2013).

\bibitem{Zhang1} Z. Y. Zhang et al., Nucl. Phys. A {\bf 625}, 59 (1997).

\bibitem{Zhang2} L. R. Dai et al., Nucl. Phys. A {\bf 727}, 321 (2003).

\bibitem{Huang2}F. Huang and Z. Y. Zhang, Phys. Rev. C {\bf 72}, 024003 (2005).

\bibitem{Huang} F.Huang et al., Chin. Phys. C{\bf 39}, 071001 (2015), and
references therein.

\bibitem{Dong}Yubing Dong, Pengnian Shen, Fei Huang, and Zongye Zhang,
Phys. Rev. C{\bf 91}, 064002 (2015).

\bibitem{PDG} K. A. Olive et al.,(Particle Data Group), Chin. Phys.
C {\bf 38}, 090001 (2014).

\bibitem{WASA1}P. Adlarson et al., ``Measurement of the $np\to np\pi^-\pi^0$
reaction in search for the recently observed $\ds(2380)$ resonance",
arXiv: 1409.2659 [nucl-ex].

\bibitem{WASA2}P. Adlarson et al., Phys. Rev. C {\bf 88}, 055208 (2013).

\bibitem{WASA3} A. Prickinga, M. Bashkanova, H. Clement, arXiv:
1310.5532,
Phys. Lett. B {\bf 727}, 438 (2013).


\bibitem{Lvqf}Qi-Fang L\"{u} et al, "Properties of single
cluster structure of $d^*(2380)$ in chiral SU(3) quark model",
arXiv:1603.03673 [hep-ph].


\end{thebibliography}
\end{document}